\documentclass[%
 preprint,
 amsmath,amssymb,
 aps,
pra,
]{revtex4-1}
\usepackage{amsmath}
\usepackage{upgreek}
\usepackage{graphicx}
\usepackage{dcolumn}
\usepackage{bm}
\usepackage{listings, physics}

\begin{document}

\title[]{High-dimension experimental tomography of a path-encoded photon quantum-state
}

\author{D. Curic}%
\author{L. Giner}%

\author{J. S. Lundeen}
\affiliation{Department of Physics and Centre for Research in Photonics, University of Ottawa, 25 Templeton Street, Ottawa, Ontario, Canada K1N 6N5}

\begin{abstract}
Quantum information protocols often rely on tomographic techniques to determine the state of the system. A popular method of encoding information is on the different paths a photon may take, for example, parallel waveguides in integrated optics. However, reconstruction of states encoded onto a large number of paths is often prohibitively resource intensive and requires complicated experimental setups. Addressing this, we present a simple method for determining the state of a photon in a superposition of $d$ paths using a rotating one-dimensional optical Fourier Transform. We establish the theory and experimentally demonstrate the technique by measuring a wide variety of six-dimensional density matrices. The average fidelity of these with the expected state is as high as 0.9852 $\pm$ 0.0008. This performance is comparable or exceeds established tomographic methods for other types of systems. 
\end{abstract}


\maketitle




\section{Introduction}

Determining the quantum state of a physical system is a key task in
quantum physics, in particular in quantum metrology,
quantum information and quantum cryptography. Since the quantum state $\pmb{\rho}$ determines all the measurable properties of system, reconstructing it can aid in building and bench-marking devices in the aforementioned areas.
Many techniques exist for reconstructing a
variety of types of physical quantum systems, ranging from electron spin to atomic position~\cite{d2003quantum,lundeen2009tomography,james2005measurement,PhysRevA.77.022307}. Considerable effort has been put towards the development of these experimental techniques, and improving their scalability. These techniques have even found applications outside their original purposes, such as in classical image processing~\cite{d1996fictitious}.

Early schemes for quantum information processing with photons proposed encoding information in a set of $d$ paths that a photon could be in~\cite{PhysRevLett.62.2124,PhysRevLett.73.58}. These paths are sometimes called rails and usually follow a waveguide or the route a classical beam would take.  The quantum state is the superposition of paths that the photon is in.  If two paths are used, the state functions as qubit. Crucially, unlike a commonly used photon qubit, polarization, higher dimensional states can be formed by adding more paths. This allows more information to be encoded on a single photon, which is then termed a `qudit'~\cite{PhysRevA.61.062308}. Higher-dimensional encodings can also simplify the design of quantum logic circuits~\cite{lanyon2009simplifying}. Path encoding is an especially popular alternative to polarization encoding for on-chip photonic devices due to the difficulty of fabricating devices that maintain polarization~\cite{matthews2009manipulation}.  A seminal contribution to manipulating path states was a scheme to implement any chosen discrete $d\times d$ unitary using an interferometric array of beamsplitters and phaseshifters acting on $d$ paths~\cite{PhysRevLett.73.58,Clements:16}.  On-chip implementations of the above-mentioned universal discrete unitaries are an especially active research direction~\cite{metcalf2013multiphoton,peruzzo2011multimode} due their ability to implement a wide set of quantum information algorithms and networks with a single device. In free-space however, paths are not a commonly used method of state encoding. This is in part due to the difficulty of reconstructing the quantum state encoded in a large number of paths. 

Despite the importance of the path-state encoding,  relatively little work has been conducted on methods to reconstruct path states with more than two dimensions. One possibility is to use a universal unitary to rotate between a complete set of incompatible bases, making projective measurements in each. This is quantum state tomography. If one universal unitary implements a quantum information algorithm another would be required for the tomography, thus doubling the complexity of the on-chip device. Moreover, in free-space, the interferometer required for the universal unitary has never been constructed due to its complexity, phase instability and alignment issues. Another possibility is to interfere each and every possible pair of paths. This has the drawback of requiring  $\mathcal{O}(d^2)$ switchable elements such as beamsplitters and mirrors. As far as we know, neither strategy for $d$-path quantum state reconstruction has been implemented above $d=2$.

On the other hand, schemes have been devised and implemented to reconstruct the full spatial quantum state of a photon~\cite{PhysRevLett.112.070405,mcalister1995optical,stoklasa2014wavefront,smith2005measurement}. The $d$-path state is embedded in this larger and continuous Hilbert space, that of the photon's transverse position. Consequently, one might consider using these techniques to reconstruct the $d$-path state.  However, these methods provide much more information than is required, for example, the modal distribution of each path. For a fixed measurement time, one would expect these methods to have a higher noise per point than a method that begins with $a$ $priori$ information that limits the size of the state-space. A method that only determines the parameters of the $d$-path density matrix $\pmb{\rho}$ would outperform these full spatial state reconstructions and may be less experimentally complicated.

We propose such a method for reconstructing a $d$-dimensional path encoded state, and implement it using both a classical and non-classical source of light. The method relies on the fact that the discrete set of states is embedded in the aforementioned continuous Hilbert space, that of the photon's transverse position. This opens up the possibility of having many more measurement outcomes than $d$. Hence, projections in many incompatible bases can be performed with one measurement apparatus configuration. This is accomplished by either measuring position $x$, with a camera, or momentum $k$, by adding a cylindrical lens that performs a one-dimensional optical Fourier transform (OFT). When the paths are distributed in both the $x$ and $y$ transverse dimensions the lens rotates to a finite set of angles in the $xy$ plane. A discrete computational Fourier transform (DFT) of the measured momentum probability distribution then allows us to easily reconstruct high-dimensional path-encoded states. In classical optics, this is equivalent to measuring the optical coherency matrix~\cite{kagalwala2015optical,PhysRevLett.62.2209}.

\begin{figure*}[ht!]
    \centering
    \includegraphics[angle=0,scale=0.3]{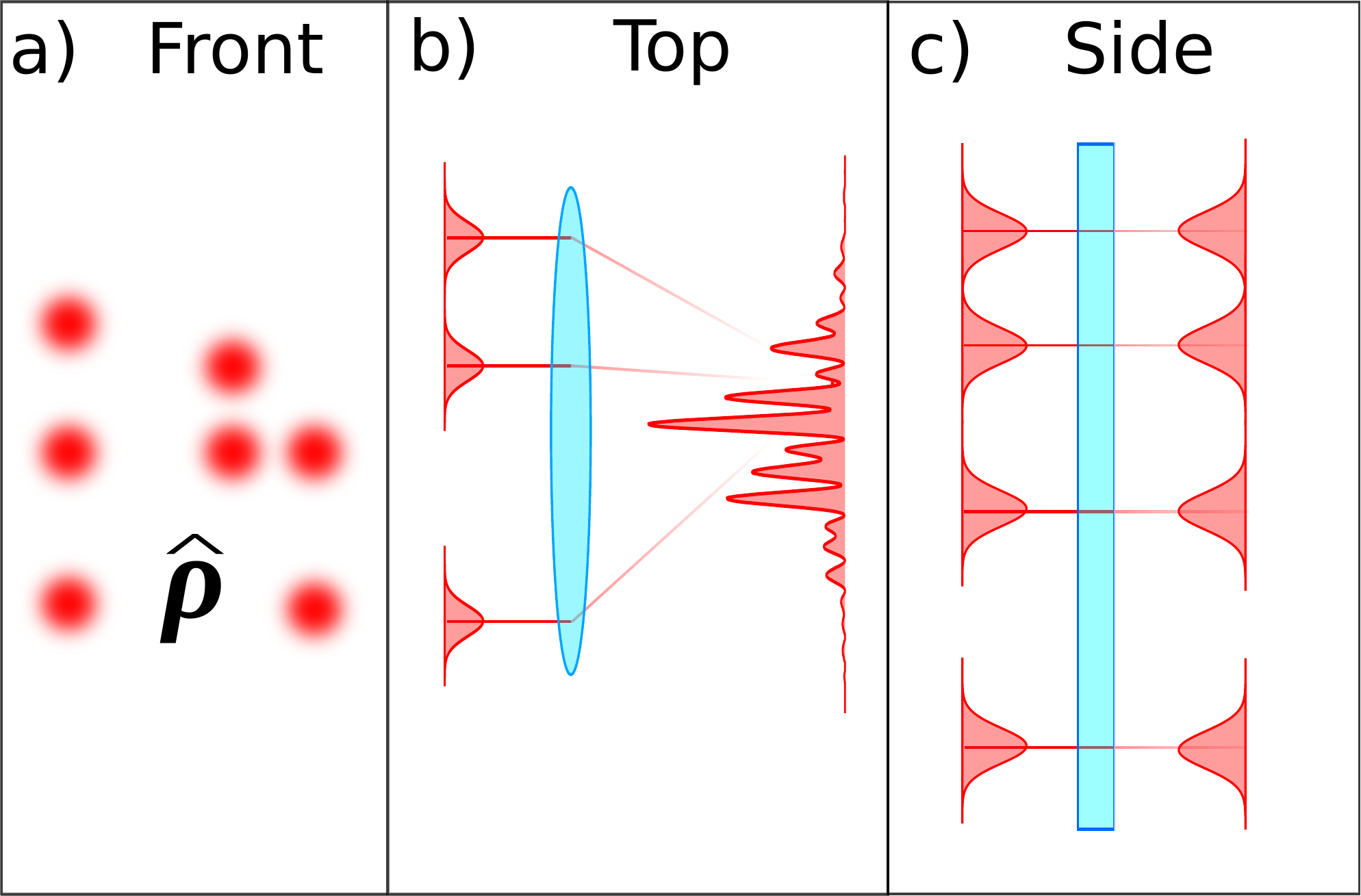}
    \caption{A conceptual diagram of the working principle of the reconstruction method. \textbf{a)} The path encoding geometry. A photon, in a superposition of being in one of the paths, will be observed in one of these paths, with probability proportional to the amplitude of that path, upon measurement. A spherical lens could not be used to reconstruct the phases and amplitudes of the paths as the spacings between paths are not unique. \textbf{b)} A cylindrical lens is used to separate the interference patterns to over come this problem. Only the interference of the three paths along the horizontal are shown for clarity. \textbf{c)} A side view of the same as the previous panel. The interference patterns are separated spatially due to the cylindrical lens.}
    \label{fig:conceptualpicture}
\end{figure*}

\section{Theory}

Before introducing our method, we begin by rigorously defining what is meant by the path-state encoding.  A single photon is travelling along \(z\) and has transverse position \(x\) (for clarity we restrict ourselves to one transverse dimension for now). A set of \(d \) states $\{\ket{\psi_i}_x\}$ are defined in this \(x\)-space. Each has a narrow position distribution $|\psi_i(x)|^2=|\braket{x}{\psi_i}_x|^2$ and is displaced in \(x\) from the other states. The displacement is much larger than the width of the any state distribution, which ensures that they have minimal overlap, $\braket{\psi_i}{\psi_j}_x\approx \delta_{ij}$ (the Kronecker delta).  Crucially, this means they form an approximate orthonormal set. In turn, this set is a basis that defines a discrete subspace embedded in the continuous \(x\) Hilbert space. Each basis state $\ket{\psi_i}_x$ represents the photon being in a particular single path $i$.  Our method additionally requires that the path-states have identical spatial distributions, other than the displacement. In the momentum basis, a displacement only affects the phase of the distribution. Hence, this identicality ensures that $ |\braket{k_x}{\psi_i}_x|^2 = |\widetilde{\psi}(k_x)|^2 $ for all \(i\), i.e. the states have identical probability distributions in momentum $k_x$. Our goal is to reconstruct states in this \(d\)-path subspace. Specifically, we will determine the density matrix $\pmb{\rho} = \sum_{ij}^d \rho_{ij} \ket{\psi_i} \bra{\psi_j}_x.$

We now describe the basic concept behind our reconstruction method.  Each diagonal element of the density $\rho_{ii}$ is the probability for the photon to be in the corresponding path $i$. These diagonals are straight-forward to determine by directly projecting onto the path states with \(d\) detectors or by measuring the position \(x\) distribution with a camera. The off-diagonal elements are more challenging. We call these elements `coherences' since a given element \(\rho_{ij}=|\rho_{ij}|\exp(\phi_{ij})\) describes the coherence between paths \(i\) and \(j\). These two paths play the same role as the slits in a double-slit interferometer. There, the coherence between the two slit-paths sets the resulting visibility (i.e. contrast) of the resulting interference pattern seen on a distant screen. Effectively, this pattern is the transverse momentum \(k_x\)  probability distribution of the photon just after the slits. Distinguishability, entanglement with other systems, amplitude imbalance, and technical or environmental noise can all decrease the visibility and, hence, the magnitude \(|\rho_{ij}|\). The phase difference between a photon passing through slit one and two, equivalent to \(\phi_{ij}\),  sets the transverse offset of the interference pattern. In summary, if one can measure the visibility and phase of the interference between the paths \(i\) and \(j\), one can determine \(\rho_{ij}\) for the path-state. 

The path states can be interfered by looking in transverse momentum (i.e. Fourier) space. Rather than using a distant screen, this can be easily accomplished with a lens of focal length $f$.  The transverse position $x_f$ of the photon $f$ distance $after$ the lens is proportional to the transverse momentum $k_x$ at $f$ distance  $before$ the lens.  The exact relationship is $x_f = f \lambda k_x/ 2\pi$, where $\lambda$ is the wavelength of the photon. In this way, a lens performs an optical Fourier transform. A general state $\rho(x,x') = \bra{x}\pmb{\rho}\ket{x'}$ in the continuous position basis corresponds to $\rho(k_x,k_x') = \bra{k_x}\pmb{\rho}\ket{k_x'}$ in the momentum basis. A camera at $f$ after the lens will thus record a signal proportional to the momentum probability distribution $P(k_x)=\rho(k_x,k_x)$. In the recorded camera distribution (i.e. image), two or more interfering paths will form an interference pattern composed of fringes.

With an illustrative two-path example, we now explicitly show how we extract \(\rho_{ij}\) from the measured momentum distribution. Our two paths, $\ket{\psi_1}_x$ and $\ket{\psi_2}_x$ are centered at $x = x_1$ and $x = x_2$, respectively. Utilizing the Fourier shift theorem, we get an explicit form for the interference pattern in terms of the path-state density matrix $P(k_x) =  |\widetilde{\psi}(k_x)|^2 \sum_{i,j}^d {\rho}_{ij} \exp(i L^x_{ij} k_x)$, where $L^x_{ij} = |x_i - x_j|$ is the pairwise distance between the paths $i$ and $j$ in the $x$ direction. Using the Hermiticity of $\pmb{\rho}$, this can be simplified to
\begin{equation}
    P(k_x) = |\widetilde{\psi}(k_x)|^2 ( \rho_{11} + \rho_{22} + 2|\rho_{12}|\cos{(L^x_{12} k_x + \phi_{12}})),
    \label{eqn:interference}
\end{equation}
where $\phi_{12}$ is the phase of $\rho_{12}$. This is the interference pattern; the fringes are due to the oscillating cosine term, and the envelope is given by Fourier Transform of a single path's spatial distribution. The visibility of an interference pattern is defined as the ratio of amplitude of the oscillation to its average value. Because $\mathrm{Tr}\{\pmb{\rho}\}=1$, the visibility can be found to be $2|\rho_{ij}|/(\rho_{11}+\rho_{22})=2|\rho_{ij}|$.  As we motivated earlier, the visibility is directly proportional to the magnitude of $\rho_{ij}$ and the phase $\phi_{ij}$ of the cosine is the phase of $\rho_{ij}$. 

We can extract the magnitude and phase information from the interference pattern in Eq.~\ref{eqn:interference} by using a discrete Fourier transform, performed with a computer. For clarity, here we present the method in terms of the continuous Fourier transform $F_{k_x}$, where the subscript denotes the transform is along the $k_x$-axis. The Fourier transform of the measured momentum distribution gives
\begin{equation} \label{eqn:interferenceFT}
\begin{gathered}
F_{k_x}\{P(k_x)\}(\bar{x}) =
  (\rho_{00} + \rho_{11})\delta(\bar{x}) + \rho_{12}\delta(\bar{x}- L^x_{12}) + \rho_{21} \delta(\bar{x} + L^x_{12}).
\end{gathered}
\end{equation}
Because this is the Fourier transform of momentum, the resulting function is back in terms of position. However, whereas a Fourier transform of the amplitude distribution would be in terms of  $x$, this is transform of the momentum \textit{probability} distribution.  We distinguish the resulting position variable by using a bar, i.e. $\bar{x}$. With this calculated distribution, it becomes simple to find $\rho_{12}$. Namely, the complex value of the Fourier transform at a position equal to spacing $L^x_{12}$ is $\rho_{12}$. That is, $\rho_{12} = F_{k_x}\{P(k_x)\}(\bar{x} = L^x_{12}).$ In short, a Fourier transform of the measured interference pattern directly gives the phase and magnitude of off-diagonal's such as $\rho_{ij}$.

While interfering paths using a lens works well for two paths, extending to higher dimensions by adding more paths introduces the possibility of overlapping peaks in $F_{k_x}\{P(k_x)\}(\bar{x})$. If more than one pair of paths have the same spacing they would contribute to the value of the Fourier transform at the same $\bar{x}$. This would make it impossible to distinguish the contributions from their corresponding density matrix elements. For instance, they may cancel completely if completely out of phase. We now discuss this problem and present our solution.


\subsection{Extension to more paths}

One way to add more path states would be to simply distribute them along \(x\). In many cases however, it may be convenient to instead distribute them in both transverse directions $x$ and $y$.  We begin by defining our path-states in two dimensions. As before, all the path states have identical position wavefunctions other than a displacement. Specifically, the $i$-th path-state $\ket{\psi_{i}}_{xy} $ is centered at position $\mathcal{P}_i = [x_i, y_i]$. We call the set of all positions $ \mathcal{P}_i $ the `geometry'.  For simplicity, we assume that each state has a position wavefunction that is a two-dimensional Gaussian with equal width $\sigma$ in both the $x$ and $y$ directions. This ensures that the distribution is symmetric under rotations in the $xy$ plane and, also, that it is a product state: $\ket{\psi_{i}}_{xy} = \ket{\psi_{i}}_x\ket{\psi_{i}}_y$.  The subscripts $x$ and $y$ distinguish the two Hilbert spaces. The position wavefunctions are, thus, $\psi_i(x,y) = \psi(x - x_i)\psi(y-y_i)$.  As before, the paths must be separated by much more than $\sigma$ so that they have minimal overlap,  $\braket{\psi_i}{\psi_j}_{x} \approx \delta_{ij}$ and $\braket{\psi_i}{\psi_j}_y \approx \delta_{ij}$. Again, this means the states form an approximate orthonormal basis for the embedded discrete Hilbert space. Our goal is to reconstruct the density matrix of the full $d$-path quantum state, $
\pmb{\rho} = \sum_{ij}^d \rho_{ij} \ket{\psi_i}_{xy}\langle \psi_{j}|$.

We must now carefully label and then distinguish every possible pair of paths since each is associated with a coherence $\rho_{ij}$. The positions of two paths, $i$ and $j$, are a connected by the line segment $\mathcal{L}_{ij}$ with end points $(\mathcal{P}_i,\mathcal{P}_j)$. The spacing of the path-pair is given by the length of the line segment, $L_{ij} = \sqrt{\left(L^x_{ij}\right)^2 + \left(L^y_{ij}\right)^2}$, where $L^{s}_{ij}=$ $|s_i-s_j|$ is the path spacing along the $s=x,y$ axis. Just as in the simple example in the last section, we obtain the coherences $\rho_{ij}$ from the recorded signal in momentum space, $P(k_x,k_y)$. However, a $d$-dimensional context allows for the possibility of \textit{multi}-path interference.  
If each line segment $\mathcal{L}_{ij}$ has a unique length and/or orientation, then each path-pair will create a signal at a distinct oscillation frequency and/or orientation in the interference pattern. In this case, all the coherences $\rho_{ij}$ can just be read out from the DFT of the two-dimensional interference pattern in $P(k_x,k_y)$. The latter could be measured by using a spherical lens to perform the optical Fourier Transform in both the $x$ and $y$ directions. The required path geometry for this, called a Golomb Rectangle, is discussed in the Appendix. However, consider if in our geometry there are two parallel line segments $\mathcal{L}_{ij}$ and $\mathcal{L}_{i'j'} $ with the same length, $L_{ij}=L_{i'j'}$. In this case, the contribution to the DFT from $\rho_{ij}$ and  $\rho_{i'j'}$ can not be distinguished since they will appear at same peak position $[\bar{x}=L^x_{ij} ,\bar{y}=L^y_{ij}]$.

\begin{figure*}[ht!]
    \centering
    \includegraphics[angle=0,scale=0.35]{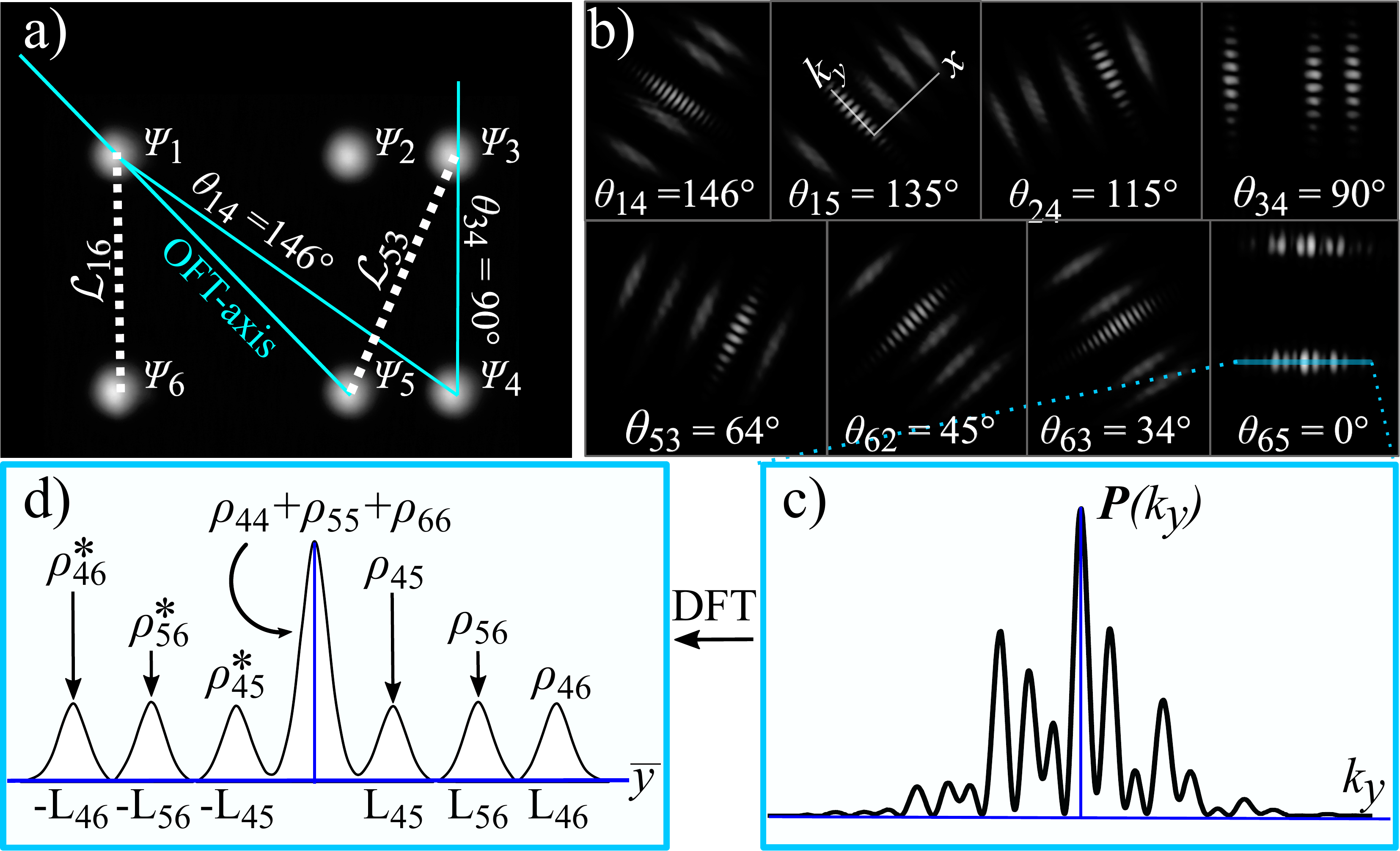}
    \caption{ State reconstruction method. \textbf{a)} The six paths are shown in the figure and encode the state $\pmb{\rho}$. The optical Fourier transform (OFT) axis (blue solid line) rotates to particular angles $\theta_{ij}$, of which a few are shown, to interfere each pair of paths at a time (angles are with respect to a horizontal axis along the bottom most paths). We assign to each pair of points a line segment $\mathcal{L}_{ij}$. \textbf{b)} The corresponding OFT for the eight angles required to reconstruct this particular density matrix. As only paths with angle $\theta_{ij}$ between them interfere, the diagonal elements can be recovered from the remaining paths. The $k_y$ axis is always in the direction of the interference and the $x$ axis is perpendicular to it (example shown for $\theta_{15}$). \textbf{c)} Each pattern is recorded and analyzed one at a time via discrete Fourier transform (DFT) by taking a one pixel wide `slice' through the interference pattern. This process is repeated for every interference pattern present in a given image. \textbf{d)} The Fourier transform of the interference pattern (For illustrative purposes we plot the magnitude). The magnitude $\rho_{ij}$ is recovered from the height of the DFT at the position $\bar{y} = L_{ij}$. The normalization is obtained by summing the zero frequency peaks of each interference pattern present in the $\theta_{65}$ subpanel (in this example) in panel $\textbf{b}$. All panels contain real data.}
    \label{fig:rotatingFTpicture}
\end{figure*}

A cylindrical lens solves this problem so long as $\mathcal{L}_{ij}$ and $\mathcal{L}_{i'j'}$ are not both segments of the same line. As an example, in Fig.~\ref{fig:rotatingFTpicture}a, $\mathcal{L}_{16}, \mathcal{L}_{25} ,$ and $\mathcal{L}_{34}$ are parallel and of equal length. As explained above, using a spherical lens will fail in this case. Consider instead the action of cylindrical lens with its OFT-axis parallel to these example line segments $\mathcal{L}_{ij}$. The resulting probability distribution will be of momentum in the direction of $\mathcal{L}_{ij}$ and position in the perpendicular direction (Fig.~\ref{fig:rotatingFTpicture}b). Critically, the line segments, $\mathcal{L}_{16}, \mathcal{L}_{25} ,$ and $\mathcal{L}_{34}$, have distinct positions along the axis perpendicular to $\mathcal{L}_{ij}$. The cylindrical lens will leave these positions unchanged. Consequently, the interference oscillation corresponding to each line segment will appear at a distinct position and, hence, will be distinguishable. In our example, we will therefore be able to measure $\rho_{16}$, $\rho_{25}$, and $\rho_{34}$ by taking a one-dimensional DFT of the recorded camera image along cylindrical lens axis. This concept can be extended to other coherences of $\pmb{\rho}$ by rotating the cylindrical lens and, hence, OFT-axis to all the unique angles $\theta_{ij}$ in our full set of line segments, where $\theta_{ij} = \tan^{-1}((y_i-y_j)/(x_i-x_j))$. An example of the recorded image for one such angle is shown in Fig.~\ref{fig:rotatingFTpicture}c-d. So long as no two $\mathcal{L}_{ij}$ and $\mathcal{L}_{i'j'}$ are segments of the same line, this procedure can be used to determine the density matrix.

For clarity in our mathematical description of this procedure, instead of rotating the OFT we rotate the full quantum state $\rho(x,x')$ in position space. Because our path wavefunctions $\psi_i(x,y)$ are rotationally invariant, this task is reduced to finding the rotated path positions $\mathcal{P}_i'$. If $R(\theta)$ is the standard rotation matrix then $\mathcal{P}_i' =  R(\theta)\mathcal{P}_i$. After this rotation, the wavefunctions are
$
\psi'_i(x,y) = \psi(x - x_i')\psi(y-y_i').
$
In the following discussion, we drop the primed notation for clarity, assuming that the paths have been rotated to the appropriate angle $\theta$. Whatever the angle, the OFT is always along the $y$-axis and the $x$-axis is left untransformed. An example of this is shown in Fig.~\ref{fig:rotatingFTpicture}b for $\theta_{34}=90^{\circ}$. The camera now records a signal proportional to probability distribution $P(x,k_y)$. 

We consider first the action of the OFT along the $y$ direction. The result is $\pmb{\rho}_x(k_y)$, which is still an operator in $x$-space, as indicated by the subscript $x$:
\begin{equation}
\begin{gathered}
\pmb{\rho}_x(k_y) = \bra{k_y} \pmb{\rho} \ket{k_y}  = 
\sum_{ij}^d \rho_{ij}  \braket{k_y}{\psi_{i}}_y \braket{\psi_{j}}{k_y} \ket{\psi_{i}} _x\bra{\psi_{j}} .
\end{gathered}
\end{equation}
Using the Fourier shift theorem, $\braket{k_y}{\psi_{i}}_y = F\{\psi(y - y_i)\} = \widetilde{\psi}(k_y)\exp(i y_i k_y)$, we simplify this to the following expression:
\begin{equation}
\begin{gathered}
\pmb{\rho}_x(k_y) =| \widetilde{\psi}(k_y)|^2\sum_{ij}^d \rho_{ij} e^{ i L^y_{ij} k_y}\ket{\psi_{i}}_x \bra{\psi_{j}}.
\label{eqn:halftransformed}
\end{gathered}
\end{equation} 
We will drop the envelope $| \widetilde{\psi}(k_y)|^2$ for the sake of brevity in the math below, but in practice it sets the width and overall height scale of the peaks in the DFT. 

To analyze the sum in Eqn.~\ref{eqn:halftransformed},  first we  separate the terms where $i = j$,
\begin{equation}
\pmb{\rho}_x(k_y) = \sum_{i}^d \rho_{ii} \ket{\psi_{i}}_x \bra{\psi_{i}} + \sum_{i \neq j}^d \rho_{ij} e^{iL^y_{ij} k_y} \ket{\psi_{i}}_x \bra{\psi_{j}}.
\end{equation}
We now find the position and momentum probability distribution $P(x,k_y)$. We evaluate it at $x=x_m$, where $x_m$ is drawn from the set of positions $\mathcal{P}_i$ of the paths in the rotated geometry. If $\ket{\psi_{i}}_x$ is not centered on $x_m$, then the wavefunction is nearly zero at $x=x_m$. In other words, $\braket{x_m}{\psi_{i}}_x = \psi(x_m-x_i)=\psi(0) \delta_{x_mx_i}$. Note that even though the values of $x_m$ may not be integers, they are from a discrete set and as such we use the Kronecker delta. With this, the probability distribution is
\begin{equation}
\begin{gathered}
 P(x_m,k_y) = \bra{ x_m} \pmb{\rho}_x(k_y) \ket{x_m} = 
 |\psi(0)|^2\left(\sum_{i}^d \rho_{ii} \delta_{x_mx_i} + 
 \sum_{i \neq j}^d \rho_{ij} e^{ i L^y_{ij} k_y} \delta_{x_mx_i}\delta_{x_mx_j}\right).
\end{gathered}
\end{equation} 
For notational convenience, we drop the $|\psi(0)|^2$ factor. In practice, this factor is accounted for when the recorded camera images are normalized.

As in our two-path example in the last section, we now take a Fourier transform of $ P(x_m,k_y)$. We do so along only the $k_y$ direction. In practice, this is implemented by performing a DFT of the camera image. The result is
\begin{equation}
\begin{gathered}
F_{k_y}\{P(x_m,k_y)\}(\bar{y}) = \int \mathrm{d}k_y P(x_m,k_y) e^{i k_y \bar{y}} = \\ \sum_{i}^d \rho_{ii}  \delta_{x_mx_i} \delta(\bar{y}) + 
 \sum_{i \neq j}^d \rho_{ij} \delta(\bar{y} -  L^y_{ij})\delta_{x_mx_i}\delta_{x_mx_j}.
 \label{eqn:finalresult}
\end{gathered}
\end{equation}
There are two cases to consider: Case 1. For a chosen $x_m$, one or more pairs of paths have $x_i=x_j=x_m$. If every pair  of these has a unique $L^y_{ij}$ then the peak at $F_{k_y}\{P(x_m,k_y)\}(\bar{y}=L^y_{ij})$ has value $\rho_{ij}$. Case 2. For a chosen $x_m$, only a single path in the rotated frame has  $x_i=x_m$. In this case,  the second term is zero and the first sum gives $F_{k_y}\{P(x_m,k_y)\}(\bar{y}=0)=\rho_{ii}$, a \textit{diagonal} element of the density matrix. Together, these two cases allow us to determine \textit{both} the diagonal and off-diagonal elements. Repeating this for the full set of cylindrical lens angles $\theta_{ij}$, the full $d$-path density matrix can be determined. This method works as long as no two parallel line segments $\mathcal{L}_{ij}$ between paths with equal spacing are along the same line, which is equivalent to the condition in Case 2. The above theoretical description constitutes the method to determine $\pmb{\rho}$ that we will later demonstrate experimentally.

\section{Experiment Setup}

\begin{figure}[ht!]
    \centering
    \includegraphics[scale=0.34]{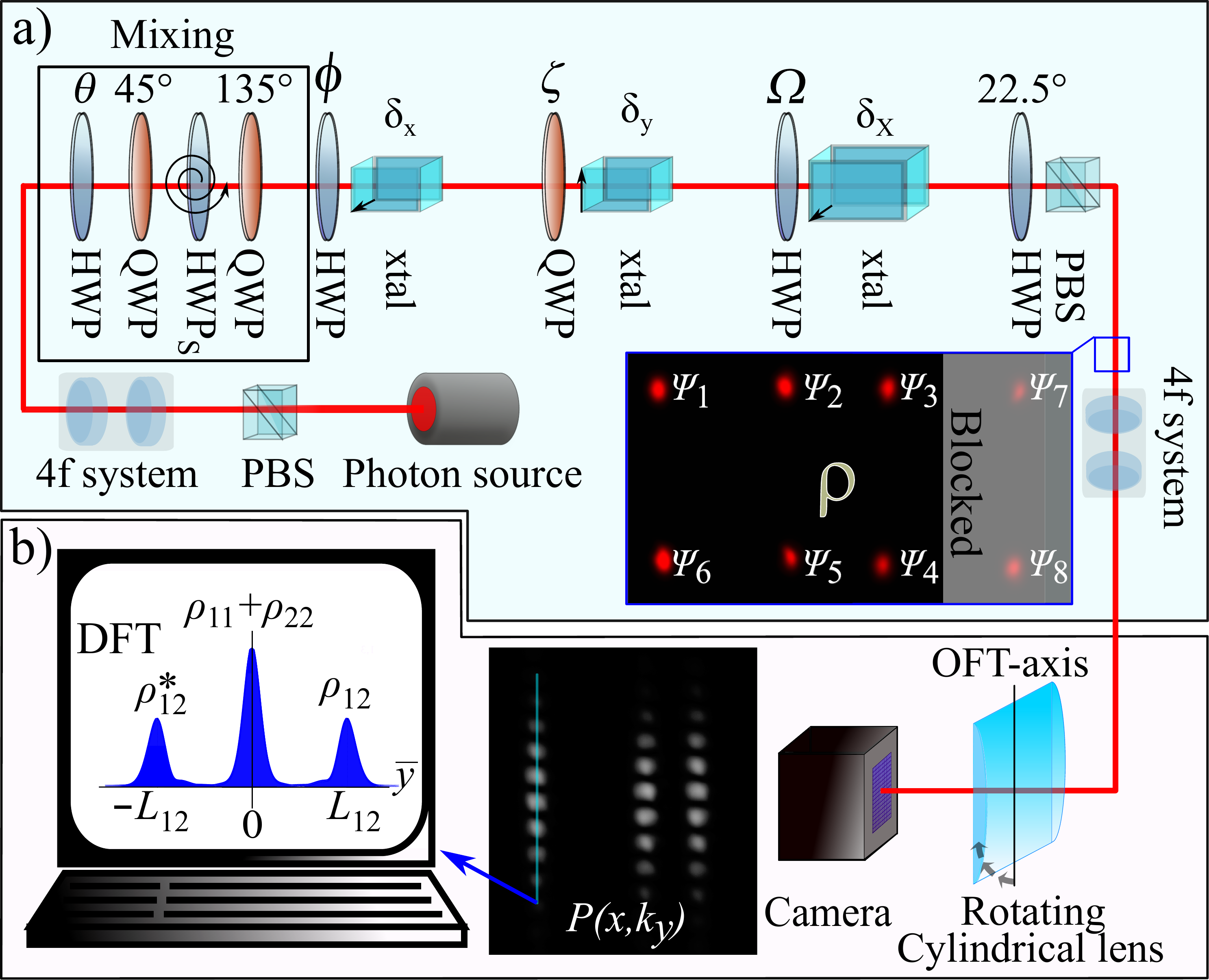}
    \caption{Experiment demonstrating the state reconstruction method. \textbf{a)} State preperation in blue box: In the blue box is the state preparation: The Rayleigh length of a $808$ nm diode laser is set by a beam expander. A series of displacement crystals (xtal) and half and quarter waveplates (labeled by the angles $\phi,\zeta$ and $\Omega$), generate the state $\pmb{\rho}$. The resulting eight path geometry is not compatible with the tomography method, and so two paths are blocked to produce a compatible six dimensional state. A set of HWP and QWP may be inserted to form a mixed state by rapidly spinning HWP$_s$. The purity is a function of the waveplate angle $\tau$. We can also produce photon pairs via SPDC at 808 nm using a diode at $\lambda=404$ nm to pump a 15 mm ppKTP crystal. The measured $g^{(2)}(0)$ of the source is  0.1979$ \pm 0.0005$.  \textbf{b)} The analysis is presented in the purple box: Lenses $f_1 = 1000$ mm and $f_2 = 400$ mm image the six paths onto a camera (a EMCCD in the case of down converted photons). A rotating cylindrical lens ($f = 250$ mm) performs the optical Fourier transform (OFT) along the OFT-axis. A one pixel wide slice of each interference pattern is analyzed with a discrete FT on a computer. Wider slices can be used and averaged over, however this may reduce the visibility if the there are imperfections in the interference pattern. This would include tilting of the dark fringes, or, as can be seen in the figure, if the intensity in each bright fringe is not evenly distributed. The coherences are obtained by the heights of the FT peaks, normalized by the total intensity.  No filters are applied to the raw data.}
    \label{fig:experimentalsetup}
\end{figure}

We demonstrate our technique by reconstructing density matrices produced by the setup shown in Fig.~\ref{fig:experimentalsetup}. Using polarized light from a diode laser at $\lambda = 808$ nm, we use a series of waveplates and calcite beam displacers, with the crystal axis cut at 45$^{\circ}$, to generate $d = 6$ paths that contain state $\pmb{\rho}$. The beam displacers utilize birefringence to split the ordinary and extraordinary polarizations of light, producing the paths, while offering robust phase stability. A 4$f$ system ensures that all optics are within the Rayleigh length of the beam. The beam waist is $340$ $\upmu$m. The first beam displacer produces a transverse shift of $\delta_x = $ 2.7 mm. The crystal axis is orientated such that horizontally polarized light is shifted horizontally. The second crystal shifts vertically polarized light vertically by $\delta_y =$ 2.7 mm as well. Lastly, the third crystal shifts horizontally polarized light by $\delta_X =$ 4 mm horizontally. In total, the beam displacers produce eight paths. However, this geometry is incompatible with the tomography method. For example, $\mathcal{L}_{13}$ and $\mathcal{L}_{57}$ are both segments of the same line and of equal length. Therefore, we block two of the eight paths to obtain a two by three path geometry which meets all the conditions necessary for the reconstruction method to work. The resulting six-dimensional density matrix is a function of the half-waveplate (HWP) and quarter-waveplate (QWP) angles $\phi,\zeta,$ and $\Omega$, preceding each crystal, shown in Fig.~\ref{fig:experimentalsetup}a. As such we write that $\pmb{\rho} = \pmb{\rho}(\phi, \zeta, \Omega)$. Note that in this situation a 2D OFT could not be used to reconstruct the state as the geometry due to repeating spacings of parallel line segments, such as, $\mathcal{L}_{12}, \mathcal{L}_{34},$ and $\mathcal{L}_{56}$ (among others), as shown in Fig.~\ref{fig:experimentalsetup}a.

The spatial distributions are imaged onto a camera using a 4$f$ system where $f_1 = 1000$ mm and $f_2 = 400$ mm, and background subtracted. The OFT is performed using a cylindrical lens ($f = 250$ mm), placed one focal length before the camera. The camera has a resolution of $3088 \times 2076$ and each pixel is square and has side-length $\gamma=2.40$ $\upmu$m. The lens rotates in an automated mount to each of the right angles $\theta_{ij}$. The one dimensional OFT produces sets of fringes, one for each pair of paths for which the angle between them is $\theta_{ij}$. To obtain the coherence, first we rotate the image back by $ - \theta_{ij}$ so as to orient the interference patterns along $y$. We then take a single line of pixels parallel to $y$ and through the center of each interference pattern. We perform a discrete Fourier transform of the recorded intensity distribution along these pixels.  Alternatively, one could average over a wider section of each interference pattern. However, averaging could potentially lower the visibility if the fringes are skewed, which could occur due to spherical aberration or misalignment of the lens.

The complex amplitude $S_{ij}$ of a peak in the DFT at spacing $\bar{y} = L^y_{ij}$  is proportional to $\rho_{ij}$. We define a scaling factor so that $|\rho_{ij}| = |S_{ij}/S|$, where $S = \sum_i^d S_{ii}$, is the total recorded intensity. We can obtain $S$ directly from the images used to calculate $\rho_{ij}$, without having to change the apparatus. Specifically, we note that in Eq.~\ref{eqn:finalresult} the following holds: $\sum_{x_m} \sum_{i}^d \rho_{ii}  \delta_{x_mx_i} =1$. In terms of the recorded signal, we get $\sum_{x_m} \sum_{i}^d S_{ii}  \delta_{x_mx_i} =S$. That is, the sum of the $\bar{y} = 0$ peaks for each $x_m$ gives the proper normalization constant $S$. This is convenient as no extra data is required to obtain this scaling.

The complex peaks also determine the phase of coherence $\rho_{ij}$ up to an overall convention and a constant offset. The offset could also be found by determining the location of $y=0$ in the camera image.  However, this can be experimentally difficult. Instead, we input a known quantum state, reconstruct it, and use it as a reference. A second reference quantum state sets our convention for which direction along the $y$ axis is positive. This is equivalent to phase-conjugation or, equivalently, which peak height $\rho_{ij}$ is proportional to in the DFT, $\bar{y} = L^y_{ij}$ or $\bar{y} = -L^y_{ij}$ . Any two states whose coherences are different and non-zero may be used as a reference.

\section{Results}

\begin{figure}[!ht]
    \centering
    \includegraphics[scale=0.38]{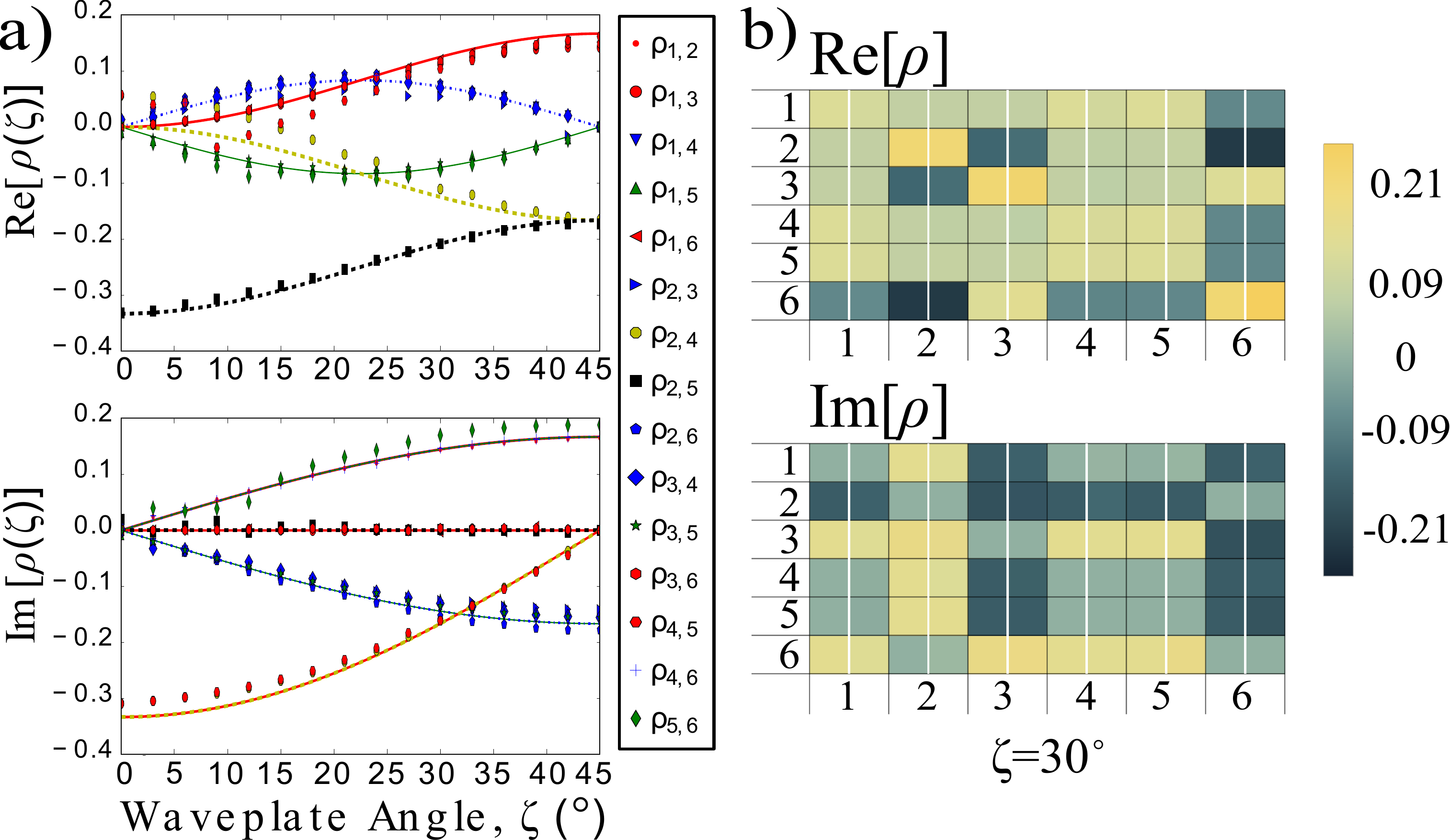}
    \caption{Experimental results. \textbf{a)} The experimental (dots) and theoretical (curve) coherences $\rho_{ij}$ of the density matrix $\pmb{\rho}$ . These are produced by varying the QWP angle $\zeta$ in Fig.~\ref{fig:experimentalsetup}. As the coherences are constrained by the experimental setup, only a few unique values appear in any given matrix. As such, data points for multiple coherences overlap. Note, error bars, obtained by averaging over multiple pictures, were omitted for clarity but range from 10$^{-3}$ to 10$^{-2}$. \textbf{b)} The experimentally reconstructed six-dimensional state $\pmb{\rho}(\zeta = 30^{\circ})$. Each diagram represents a 6$\times$6 matrix, with theoretical elements to the right of each experimental element.The fidelity with the nominal input state is 0.9911 (fidelity is one if the states are identical).}
  
    \label{fig:cohasfuncofwp}
\end{figure}

The reconstruction method is demonstrated with $d=6$ dimensional states of the form $\pmb{\rho}(\phi = 22.5^{\circ},\zeta,\Omega = 22.5^{\circ}) = \pmb{\rho}(\zeta)$ (i.e., all HWP are held fixed), whose coherences are a function of the QWP angle $\zeta$. Fig.~\ref{fig:cohasfuncofwp}a shows the real and imaginary parts of the fifteen off-diagonal elements $\rho_{ij}$ (not including the conjugates, $\rho_{ij}$). Note that,  several elements of $\rho_{ij}$ will be equal since the set of states the experimental setup can produce is constrained. We see good agreement between the experimental values (dots) and the theoretical values (dotted lines). Fig.~\ref{fig:cohasfuncofwp}b shows the experimental and theoretical density matrix for $\zeta = 30^{\circ}$ in a side by side comparison.

\begin{figure}[!ht]
    \centering
    \includegraphics[scale=0.38]{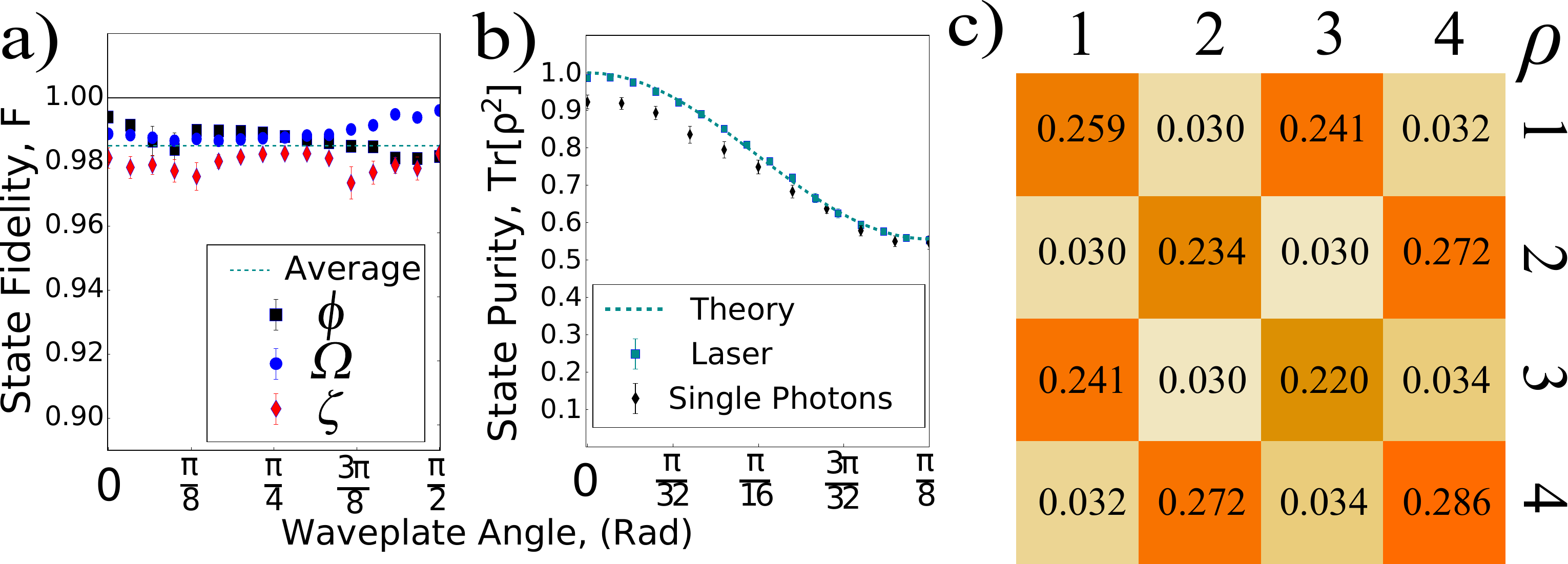}
    \caption{Experimental results. \textbf{a)} The fidelity as a function of the waveplate angles $\phi$, $\zeta$ and $\Omega$, shown in Fig.~\ref{fig:experimentalsetup}. The fidelity is close to unity, meaning $\pmb{\rho}$ and $\pmb{\rho}_{\mathrm{th}}$ are nearly equal. Averaged over all points the fidelity is $0.9852\pm0.0008$ (dashed line). \textbf{b)} The purity Tr$\{\pmb{\rho}^2\}$ as a function of the angle of the HWP angle ${\tau}$ for a classical source and single photon source. The single photon source deviates due to the much shorter coherence length. \textbf{c)} The reconstructed density matrix of single photons in four paths. The experimental values are labeled. The corresponding theoretical values are either 0.25 or 0. The calculated fidelity is 1.00 $\pm 0.03$.}
    \label{fig:fidpur}
\end{figure}

Next, we calculate the state fidelity, $F(\pmb{\rho},\pmb{\sigma}) = \mathrm{Tr}\{\sqrt{\sqrt{\pmb{\rho}} \pmb{\sigma} \sqrt{\pmb{\rho}}}\}$, which is a measure of how close two density matrices $\pmb{\rho}$, and $\pmb{\sigma}$ are. The fidelity is bound between zero and unity, and $F(\pmb{\rho},\pmb{\sigma}) = 1$ only if $\pmb{\rho} = \pmb{\sigma}$. We calculate the fidelity between the experimentally reconstructed state $\pmb{\rho}(\phi,\zeta,\Omega)$ and the theoretically predicted one $\pmb{\rho}_{\mathrm{th}}$. This is done for sets of states where one of the waveplate angles ${\phi}$, ${\Omega}$, or ${\zeta}$ are varied. Specifically, if a HWP is not being varied, then it is fixed at 22.5$^{\circ}$. If either HWPs ${\phi}$ or ${\Omega}$ are varied then we fix $\zeta = 45^{\circ}$. Fig.~\ref{fig:fidpur}a displays the results. The average fidelity when ${\phi}$ is varied is $0.987\pm 0.001$, $0.9893\pm 0.0008$ when ${\Omega}$ is varied, and $0.979 \pm 0.001$ when ${\zeta}$ is varied. Average over all trials the fidelity is $0.9852\pm0.0008$. Given the fidelity is consistently close to one, we conclude that the reconstructed state faithfully reconstructs the theoretical state.

The density matrix generalizes the concept of the state by accommodating statistical mixtures of states. A measure of this is the state purity Tr$\{\pmb{\rho}^2\}$. Pure states are those for which the purity is one. Statistical mixtures decrease the purity to a minimum of $1/d$, where $d$ is the dimension of the Hilbert space. We reconstruct $\pmb{\rho}$ and calculate the purity. To mix the state we introduce a series of HWP-QWP-HWP$_s$-QWP, shown in Fig.~\ref{fig:cohasfuncofwp}a. Here the QWP are 90$^{\circ}$ apart. The subscript $s$ on HWP$_s$ indicates that it is rapidly spinning in a mount. HWP$_s$ spins faster than the collection time of the camera. This mixes the polarizations, which in turn mixes the path state density matrix. The first HWP controls the degree of mixing. The waveplate angle $\tau = 0^{\circ}$ produces a pure state and $\tau =22.5^{\circ}$ produces a totally mixed polarization state. The theoretical purity of the resulting path state $\pmb{\rho}(\tau)$ is Tr$\{\pmb{\rho}^2(\tau)\} = 1/9(5 + 4 \cos^2{(4 \tau)})$. Fig.~\ref{fig:fidpur}b shows that the measured purity agrees strongly with the theory. 

To demonstrate that the technique is applicable to non-classical sources of light we switch to a source of down converted photons. Along with the significantly lowered average photon count number, single photon sources have the added challenge that the coherence length is also substantially reduced. To accommodate for this fact, we introduce glass slides in the paths of the photons to compensate for the path length difference imparted by the displacement crystals. We pump a 15 mm ppKTP crystal with a 404 nm laser to produce degenerate pairs at 808 nm. The measured autocorrelation $g^{(2)}$ of the source is 0.1979 $ \pm\; 0.0005$, which is much less than one thereby confirming the light is antibunched and non-classical. We generate a 4$\times$4 path encoded density matrix in a square geometry with  each side length being 2.7 mm. We replace the QWP labeled by $\zeta$ with a HWP, and set all HWP at 22.5$^{\circ}$. The purity of this matrix also depends on $\tau$. The experiment is otherwise the same, except that the down-converted photons are imaged onto an electron multiplying CCD (EMCCD). The EMCCD has a resolution of $512 \times 512$ with a pixel length of $\gamma = 16$ $\upmu$m. The calculated purity is shown in Fig.~\ref{fig:fidpur}b. Because the path compensation is not perfect, some elements have decohered, leading to reduced measured purity. Nevertheless, the results follow the theoretical trend well. In Fig.~\ref{fig:fidpur}c we plot the density matrix elements for $\tau = 22.5^{\circ}$, and list the experimental values. For reference, every theoretical elemlent is either 0.25 or 0. The calculated fidelity is  1.00 $\pm 0.03$, which confirms that the method works well for quantum light sources such as single photons. 

The reconstruction method satisfies the unity trace and hermiticity conditions, but, like in most linear reconstruction methods, the state may not be positive
semi-definite, as is required by a physically possible density matrix. Positive semi-definitivity requires that $|\rho_{ij}| \le \sqrt{\rho_{ii} \rho_{jj}}.$ Indeed, experimental imperfections cause calculated fidelity of the single photon density matrix, 1.00 $\pm 0.03$, attains values greater than one when the uncertainty is added. Imposing positive semi-definitivity would require maximum-likelihood algorithms to fit the most likely physical state~\cite{baumgratz2013scalable}. Nevertheless, we see that even without these likelihood algorithms, metrics such as fidelity and purity, are still very high. In the Appendix, we discuss the experimental resource requirements of the present tomographic method, as well explicit algorithms for constructing compatible geometries of path states.

\section{Summary}

The above results show the reconstruction method faithfully reconstructs several different metrics for a six dimensional density matrix. The reconstruction method exploits the fact that discrete paths are embedded in a larger, continuous space. This allows us to use Fourier transforms in a continuous space to obtain data for a discrete density matrix. To avoid having multiple signals with the same spacing overlap, we utilize a cylindrical lens to separate the interference patterns spatially along the orthogonal direction. In this sense, the method uses both position space and $k$-space simultaneously to reconstruct the state. We emphasize that the above results reflect a proof of principle; the tomography method presented above can, in theory, accommodate much larger dimensions, given they conform to the physical constraints (e.g. pixel size, camera size, aberration etc.) discussed above. We demonstrate the method experimentally by calculating several metrics for six dimensional matrices. In summary, we present an experimentally simple method using one dimensional Fourier transforms, one optical and the other digital, to reconstruct large dimensional density matrices encoded in the paths of photons.

\section*{Acknowledgements}
This work was supported by the Canada Research Chairs (CRC) Program, the Canada First Research Excellence Fund (CFREF), and the Natural Sciences and Engineering Research Council (NSERC).

\section*{Appendix}

\subsection*{Counting the resources required for the method}

We briefly consider the resources needed for the method. This allows one to compare it to other quantum-state reconstruction methods such as tomography. We begin with the required number of experimental configurations $\eta$, which are sometimes called "measurement settings". In our method, the only experimental change from one measurement to the next is the OFT-axis angle. The number $\eta$ of angles $\theta_{ij}$ required to determine all the density matrix off-diagonal elements $\rho_{ij}$ depends on the specific geometry. In the worse case, each lens angle only retrieves one element. Thus the maximum number of angles would be $d(d - 1)/2$, the number of independent off-diagonal complex elements in $\pmb{\rho}$. More typically, a number of paths will lay along the same line $\mathcal{L}$. In this case, multiple off-diagonals can be found with a single lens angle aligned along $\mathcal{L}$. Consequently, judicious choice of geometry can reduce the total number of measurement settings significantly. 

Consider a one-dimensional geometry. That is, the path-states are arranged along, say, the $y$-axis. In this case, only one experimental setting is needed to determine the all the density matrix off-diagonals.  The lens axis is set parallel to the line of path-states along $y$. The diagonals can be found with one additional setting, the lens axis along $x$. Valid one-dimensional geometries are given by a "Golomb ruler", a set of points with no repeated spacings. An optimal Golomb ruler is one that minimizes the maximum spacing for a number of points, $d$. In the limit of large $d$, the Er\"os-Turan construction for an asymptotically optimal set is $x_{i+1}=L_{\mathrm{min}}(2di+(i^2\,\bmod\,d)),$ for $i=0,2,...d-1$ for  any value of $d$ that is and odd prime~\cite{erdos1941problem}. Here, $L_{\mathrm{min}}$ is the minimum possible spacing between paths. The maximum spacing, $L_{\mathrm{max}}=x_d-x_1$, is then given by $L_{\mathrm{max}} =L_{\mathrm{min}}(2d(d-1)+1)$.  Space is a resource for both the path-state encoding and our reconstruction method.  The space available (e.g. the size of the camera screen) constrains the maximum possible number of path-states. Considering a one-dimensional arrangement, this maximum is $d_{\mathrm{max}}=L_{\mathrm{max}}/L_{\mathrm{min}}$.  In contrast, a Golomb ruler uses space less efficiently: $d=\sqrt{L_{\mathrm{max}}/(2L_{\mathrm{min}})}=\sqrt{d_{\mathrm{max}}/2}$ for large $d$.  A one dimensional geometry is optimal for the number of measurement settings required to reconstruct $\pmb{\rho}$ (i.e., $\eta = 2$), but sub-optimal in its use of space. 

We now examine whether a two-dimensional geometry uses space more efficiently. Robinson gave a generalization of the Golomb ruler called the Golomb rectangle~\cite{robinson1979optimum}. In the area of astronomical inteferometers, these are known as "non-redundant configurations"~\cite{kopilovich2013construction}. In it, every line segment $\mathcal{L}_{ij}$ is associated with a unique vector, i.e. orientation and length. Given a non-redudant configuration, a spherical lens is sufficient to reconstruct $\pmb{\rho}$. As with the one-dimensional geometry, two measurement settings are sufficient: lens present and lens absent. Given $L_{\mathrm{max}}$ and $L_{\mathrm{min}}$, explicit computational searches for optimal Golomb rectangles suggest that maximum number of path-states is $d=\mathcal{O}(( L_{\mathrm{max}}/L_{\mathrm{min}}))$~\cite{kopilovich2013construction}. In two dimensions,  the maximum number of closely packed path-states is $d_{\mathrm{max}}=\mathcal{O}((L_{\mathrm{max}}/L_{\mathrm{min}})^2)$. Consequently, the optimum Golomb rectangles achieve $d=\mathcal{O}(\sqrt{d_{\mathrm{max}}})$. In terms of the total dimension of the discrete path-state Hilbert space, the benefit of going from one to two dimensions is limited to the an overall constant factor. 

A number of experimental factors will constrain the largest feasible $L_{\mathrm{max}}$. The most obvious is the range of positions $D$ over which OFT lens system correctly performs a Fourier Transform. This will be set by lens aberrations and/or the lens diameter.  Less obvious is the pixel pitch of the camera $\gamma$, the distance between pixel centers. The finest interference pattern produced by the paths will have a fringe spacing of  $2 \pi L_{\mathrm{max}}/ \lambda f$ at the camera, where $f$ is the focal length of the OFT lens. In order to sample the fringes without aliasing, the Nyquist condition then constrains $L_{\mathrm{max}} < \lambda f / \pi \gamma$. On the other hand, $L_{\mathrm{min}}$ is constrained primarily by the width of the paths.

Another resource to consider is the number of measurement outcomes $N$ per measurement setting. In our case, this is the number of camera pixels. As described above, the finest pixel pitch required is $\gamma = \lambda f / \pi L_{\mathrm{max}}$ . However, the camera pixel-array width $W$ must be large enough to record the largest fringe spacing, which in turn, results from the smallest spacing $L_{\mathrm{min}}$. This sets $W=\lambda f / \pi L_{\mathrm{min}}$. Together, these requirements determine the number of pixels along one camera dimension, must satisfy $N>W/\gamma=L_{\mathrm{max}}/L_{\mathrm{min}}=d_{\mathrm{max}}$. As discussed above $L_{\mathrm{max}}<D$.  In addition, $L_{\mathrm{min}}$ must be greater than the path-state position-width, $L_{\mathrm{min}}>\sigma$.  Taken together, we have the requirement $N>D/\sigma$. Or, using the results from the optimal Golomb ruler above, $N>2d^2$ for large $d$. In typical quantum state tomography, $N=d$. These arguments can be straightforwardly extended to two dimensions. This shows that, not only is our method sub-optimal in its use of space, but also in terms of the number of measurement outcomes per measurement setting. Of course, to its credit, it is, so far, the only experimentally feasible method for path-state reconstruction and it only requires two settings.

Perhaps these resources could be traded for one another by using a cylindrical lens. The use of a cylindrical lens permits a denser usage of space since, as long as they are not collinear, two or more line segments may have identical vectors. It follows that the ratio of $d$ to $N$ and $d_{\mathrm{max}}$ will improve. The trade-off is that multiple lens angles and, thus, measurement settings $\eta$ , will be necessary.  As far as we know, the required path-state geometry has not been studied and, so, optimal solutions do not exist. Consequently, we are unable to evaluate this resource trade-off in more detail.

\bibliographystyle{apsrev4-1}
\bibliography{sample}

\end{document}